\documentclass[a4paper,11pt]{article}
\usepackage{amsfonts}

\usepackage{graphicx}
\usepackage{amsmath}
\usepackage{amssymb}
\usepackage{latexsym}
\usepackage[colorlinks]{hyperref}
\usepackage{color}
\usepackage{float}
\usepackage{cite}
\usepackage{makeidx}
\usepackage{colortbl}
\usepackage{csquotes}
\usepackage[squaren]{SIunits}

\usepackage[textfont={footnotesize,sf},labelfont={color=blue,bf,sf},labelsep=endash]{caption}
\usepackage[position=top,labelfont={color=blue,bf,sf}]{subfig}
\usepackage{bm}

\usepackage[title]{appendix}

\newcommand{\bra}{\begin{array}}
    \newcommand{\era}{\end{array}}
\newcommand{\beq}{\begin{equation}}
\newcommand{\eeq}{\end{equation}}
\newcommand{\bqr}{\begin{eqnarray}}
\newcommand{\eqr}{\end{eqnarray}}

\def\BC{\bb C}
\def\_\BC{\bbi C}

\def\no2 {{\textstyle{n\over 2}}}

\newcommand{\om}{\omega}

\newcommand{\al}{\alpha}

\newcommand{\lb}{\label}

\begin{document}
    \begin{titlepage}
        \setcounter{page}{1}
        \renewcommand{\thefootnote}{\fnsymbol{footnote}}

        \begin{flushright}
        \end{flushright}

        \vspace{5mm}
        \begin{center}

{\Large \bf {
Goos-H\"anchen Shifts in Gapped
Graphene\\ subject to External Fields}}

\vspace{5mm}

 {\bf Miloud Mekkaoui}$^{a}$, {\bf Ahmed Jellal\footnote{\sf a.jellal@ucd.ac.ma}}$^{a,b}$
 and {\bf Hocine Bahlouli}$^{c}$

\vspace{5mm}

{$^{a}$\em Laboratory of Theoretical Physics,  
Faculty of Sciences, Choua\"ib Doukkali University},\\
{\em PO Box 20, 24000 El Jadida, Morocco}

{$^{b}$\em Canadian Quantum  Research Center,
204-3002 32 Ave Vernon, \\ BC V1T 2L7,  Canada}

{$^c$\em Physics Department, King Fahd University of Petroleum $\&$ Minerals,\\
 Dhahran 31261, Saudi Arabia}

\vspace{3cm}

\begin{abstract}
We study Dirac fermions in gapped graphene that are subjected to a magnetic field  and a
potential barrier  harmonically oscillating in time. The tunneling modes inside the gap and the
associated  Goos-Hänchen (GH) shifts are  analytically investigated. We show
that the GH shifts in transmission for the central band and the first two sidebands change sign at the Dirac
points  $\epsilon+l\hbar\tilde{\omega}$ $(l=0,\pm 1)$.
We also find that the GH shifts can be
either negative or positive and becomes zero at transmission resonances.

\vspace{3cm}



\end{abstract}
\end{center}
\end{titlepage}


\section{ Introduction}

Investigations of transport properties in periodically driven quantum systems is not only of academic interest
but also is of great importance in the design of novel devices and optical applications. In particular, quantum interference
within an oscillating time-periodic electromagnetic field gives
rise to additional frequency contributions $\epsilon \pm l\hbar \tilde\omega$
$(l=0,\pm 1, \cdots)$ in the transmission probability. 
These can be interpreted to originate from exchanging energy quanta $\hbar\tilde \omega $ with the oscillating field. Hereby $\tilde\omega $ is the oscillation frequency and for electromagnetic waves the quanta, that are
exchanged with the electrons, are photons.  In this context, the standard model  is that
of a time-modulated scalar potential in a finite region of space.
It was studied earlier by Dayem and Martin \cite{Dayem} who
provided the experimental evidence of photon assisted tunneling in
experiments on superconducting films under microwave fields but
 Tien and Gordon \cite{Tien} provided the first theoretical
explanation of their discovery. 
Afterwards, further
theoretical studies were performed  by many research groups,
for instance 
the barrier
traversal time of particles interacting with a time-oscillating
barrier was investigated in \cite{Buttiker, Mekkaoui2014}. Also 
the treatment on photon-assisted transport through
quantum wells and barriers with oscillating potentials by analyzing 
in depth the transmission probability as a function of the
potential parameters were done in \cite{Grossmann,Wagner}.

 In the past few years  the
optical properties in graphene systems such as the quantum version
of the Goos-H\"anchen (GH) effect originating from the reflection
of particles from interfaces have been studied. The GH effect was discovered by
Hermann Fritz Gustav Goos and Hilda H\"anchen \cite{Goos} and
theoretically explained by Artman \cite{Artmann} in the late of
1940s. Studies of  various graphene-based nanostructures,
including single \cite{xChen1}, double barrier \cite{Song} and
superlattices \cite{xChen2, Kamal}, showed that the GH shifts can be
enhanced by the transmission resonances and controlled by varying
the electrostatic potential and induced gap. Similar
to observations of GH shifts in semiconductors, the GH shifts in
graphene can also be modulated by electric and magnetic barriers
\cite{Sharma}, an analogous GH like shifts can also be observed in
atomic optics \cite{Huang}. It has been reported that the GH shifts
play an important role in the group velocity of quasiparticles
along interfaces of graphene p-n junctions \cite{Beenakker, Zhao}.
Experimentally,
 it was observed that depositing
 graphene  on dielectric materials
can result in a profound effect on GH shifts, which can be either
positive or negative. Strikingly this approach
allows complete electrostatic control
\cite{Jiang1, Jiang2}. Recently it has been shown that nonlinear
surface plasmon resonance in graphene can provide rigorous
enhancement and control over GH effect \cite{You1, You2}.

 We generalize the results obtained in our previous work \cite{Mekkaoui} to
include a
 magnetic field case by studying graphene sheet lying in the $xy$-plane that is subjected   to a scalar square potential barrier
 along the
$x$-direction while the carriers are free in the $y$-direction.
The barrier height oscillates sinusoidally around an average value
$V$ with oscillation amplitude $U_1$ and frequency $\omega$. 
The solutions of the energy spectrum are obtained for all modes generated
by the oscillating potential.
The boundary conditions  are applied at interface  to explicitly determine the associated  transmission
probabilities.
We
calculate the GH shifts in transmission for the central band and
the side sidebands as a function of the potential parameters, 
incident angle of the particles and phase shifts. 
In particular,
we show that GH shifts in transmission can be
controlled by a magnetic barrier driven by the time-periodic
scalar square potential.

 The manuscript is organized as
follows. In section $2$, we formulate our theoretical model by setting the
Hamiltonian system describing particles scattered by a single
barrier time-oscillating whose intermediate zone is subject to a magnetic field and
mass term. We determine the  quasi-energy spectrum and the spinor solution corresponding to each region
composing our system. 
In
section $3$, we use the solutions 
associated
with our system together with transmission probabilities to
compute the GH shifts. 
To
acquire a better understanding of our results, we plot the GH
shifts in transmission within the central band and the first two sidebands
for different values under suitable conditions
in section 4. Our conclusions
are given in the final section.

\section{ Theoretical model}

\begin{figure}[H]
        \centering
        \subfloat[]{
            \centering
            \includegraphics[height=4.0cm]{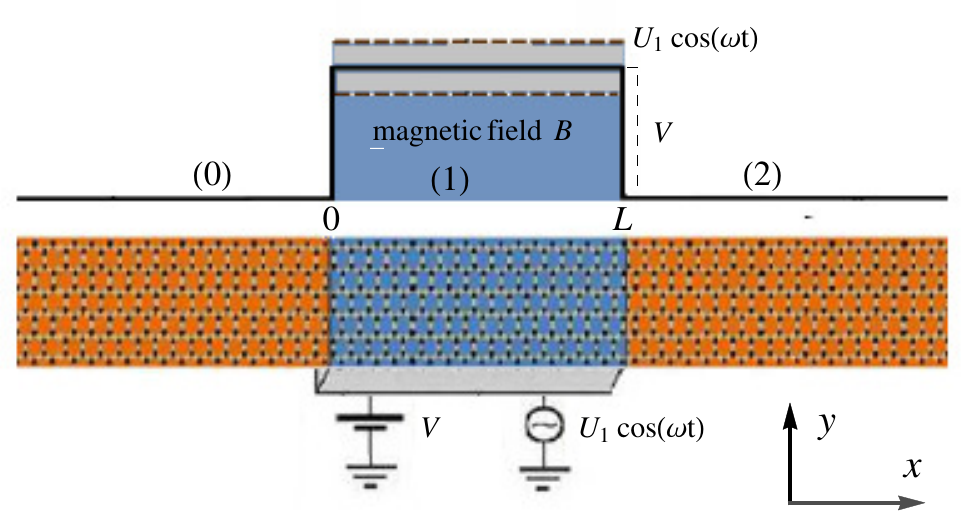}
            \label{db.3}
        }\subfloat[]{
            \centering
            \includegraphics[height=4.0cm]{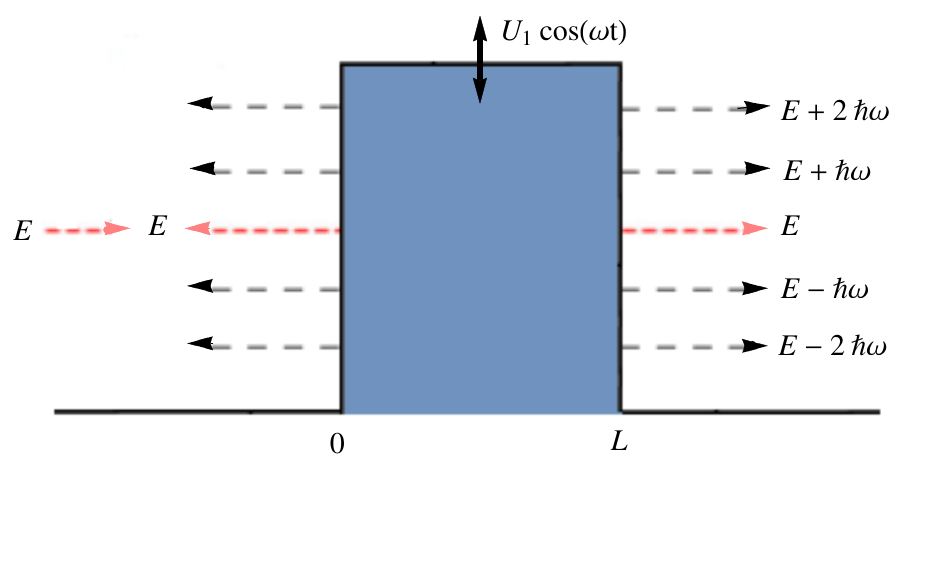}
            \label{db.4}}
    \caption{\sf{(color online) {\color{blue}(a)}: Profile of a magnetic barrier driven by a periodic
potential with frequency $\omega$, amplitude $U_1$, width $L$ and height $V$ applied
   to graphene composed
of three regions.
    {\color{blue}(b)}: Schematic of potential barrier oscillating in time. }}
    \label{fig1}
\end{figure}

We consider a flat sheet of graphene in the presence of 
a square potential barrier along the $x$-direction while particles
are unrestricted in the $y$-direction. The width of the barrier is $L$,
its height is oscillating sinusoidally around $V$ with amplitude
$U_1$ and frequency $\omega$. The intermediate zone is subject to
a magnetic field perpendicular 
$\textbf{B}=B(x, y)\textbf{e}_z$ and mass term $\Delta$. Particles with energy $E=v_F
\epsilon$ are incident from one side of the barrier at an angle
$\phi_{0}$ with respect to the $x$-direction. They leave the
barrier with energy $\epsilon+ l\hbar \tilde\omega$, with $l=0, \pm 1,
\cdots$ are the modes generated by the oscillating potential at frequency $\omega= v_F \tilde\omega$,  and they
make angles $\pi-\phi_{l}$ in the reflection and  $\theta_{l}$ in transmission regions. Our  system is governed by  the  Hamiltonian 
\begin{equation}\lb{ham1}
H=H_{\sf I}+H_{\sf II}
\end{equation}
such that $H_{\sf I}$ is given by
\begin{equation}\lb{ham2}
H_{\sf I}=v_{F} {\boldsymbol{\sigma}} \cdot \left(-i\hbar
{\boldsymbol{\nabla}}+
\frac{e}{c}\textbf{A}(x,y)\right)+V(x){\mathbb
I}_{2}+\Delta\Theta(Lx-x^{2})\sigma_z
\end{equation}
and $H_{\sf II}$ describes the harmonic time dependence of the
barrier height
\begin{equation}
H_{\sf II}=U_{j}\cos(\omega t) \mathbb{I}_2
\end{equation}
where $\upsilon_{F}$ is the Fermi velocity, $
{\boldsymbol{\sigma}} =(\sigma_{x}, \sigma_{y})$ are the  Pauli
matrices,  ${\mathbb I}_{2}$ is the $2 \times 2$ unit matrix,
 the amplitudes of static square potential barrier $V$ and
of  oscillating potential $U_j$ are given by
\begin{equation}
V(x)=
\left\{%
\begin{array}{ll}
    V, & \qquad\hbox{$0\leq x\leq L$} \\
    0, & \qquad \hbox{otherwise} \\
\end{array}%
\right.,\qquad U_{j}=
\left\{%
\begin{array}{ll}
    U_1, & \qquad \hbox{$0\leq x\leq L$} \\
    0, & \qquad \hbox{otherwise} \\
\end{array}%
\right.
\end{equation}
and the script $j = {\sf 0}, {\sf 1}, {\sf 2}$ denotes each
scattering region. For a magnetic barrier, the relevant physics is
described by a magnetic field translationally invariant along the
$y$-direction, $B(x, y)= B(x)$
within the
strip $0\leq x\leq L$ but $B=0$ elsewhere, which can be formulated
in terms of the Heaviside step function $\Theta (x)$ as
\begin{equation}
B(x,y)= B_{0}\Theta(Lx-x^{2}).
\end{equation}
Choosing the Landau gauge
imposes the vector potential  ${\boldsymbol{A}} =
(0,A_{y}(x))^{T}$ with $\partial_{x}A_{y}(x)= B(x)$ and thus the
transverse momentum $p_{y}$ is conserved.
The continuity of ${\boldsymbol{A}}$ requires that
\begin{equation}
\qquad A_{y}(x)=\left\{%
\begin{array}{ll}
    0, & \qquad \hbox{$x<0$} \\
    B_{0}x, & \qquad \hbox{$0\leq x\leq L$} \\
    B_{0}L, & \qquad \hbox{$x>L$}. \\
\end{array}%
\right.
\end{equation}
Strictly speaking,
our system can be visualized as in Figure \text{\ref{fig1}}{\color{blue}a}, which shows clearly  the regions of oscillating potential
and magnetic field. In  Figure \text{\ref{fig1}}{\color{blue}b}, we picture the energy modulations that are due the
oscillating potential. This will help us to analyze the tunnel effect
and calculate different physical transport quantities.

Based on the results presented in {\bf Appendix} \ref{AppendA},
we summarize our main solutions by writing the scattering
states in different regions. Recall that, in regions  {\sf 0}
and  {\sf 2} the potential height is $u_j=0$, then we proceed by
replacing $J_{m-l}$ by $\delta_{ml}$. Consequently,  we have
in region
{\sf 0} ($x<0$)
\begin{equation}
\psi_{\sf
0}(x,y,t)=e^{ik_{y}y}\sum^{+\infty}_{m,l=-\infty}\left[\delta_{l0}
\left(
\begin{array}{c}
1 \\
 \alpha_{l}\end{array}\right)e^{ik_{l}x}+r_{l}\left(
\begin{array}{c}
1 \\
 -\frac{1}{\alpha_{l}}\end{array}\right)e^{-ik_{l} x
 }\right]\delta_{ml}\  e^{-iv_{F}(\epsilon+m\tilde{\omega})t}
\end{equation}
and in region {\sf 2} $(x>L)$
\begin{equation}
\psi_{\sf
2}(x,y,t)=e^{ik_{y}y}\sum^{+\infty}_{m,l=-\infty}\left[t_{l}\left(
\begin{array}{c}
1 \\
 \beta_{l}\end{array}\right)e^{ik^{'}_{l} x
 }+b_{l}\left(
\begin{array}{c}
1 \\
 -\frac{1}{\beta_{l}}\end{array}\right)e^{-ik^{'}_{l} x}\right]\delta_{ml}\
  e^{-iv_{F}(\epsilon+m\tilde{\omega})t}
\end{equation}
where $\{b_{l}\}$ is a set of the null vectors, $\{r_{l}\}$ and  $\{t_{l}\}$
are the reflection and transmission coefficients, respectively.

As far as region {\sf 1} ($0\leq x \leq L$)  is concerned, where
$H_{\sf II}$ is non zero, and according to \cite{Tien}
the eigenspinors $\psi_{\sf 1}(x,y,t)$ of
the total Hamiltonian  \eqref{ham1} can be expressed in terms of the
eigenspinors $\psi_{1}(x,y)$ at energy $\epsilon$ of $H_{\sf I}$. Then, we have
\begin{equation}\lb{state1}
{\psi_{\sf 1}(x,y,t)=\psi_{1}(x,y)\sum^{+\infty}_{m=-\infty} J_{m}
\left(\frac{u_j}{\tilde{\omega}}\right)\ e^{-iv_{F}(\epsilon+\tilde{\omega} m)t}}
\end{equation}
where $J_{m}\left(\frac{u_j}{\tilde{\omega}}\right)$ are Bessel functions.
We should emphasis that \eqref{state1} is found by
resolving the Schr\"odinger equation and requiring that the function
$f=\sum_m J_m\left(\frac{u_j}{\tilde{\omega}}\right) e^{-im\omega t}$ must fulfill the condition $f'=H_{\sf II} f$.
To include all modes, a linear combination of spinors at energies
$\epsilon+l\tilde{\omega}$ $(l=0,\pm 1,\cdots)$ has to be taken.
Hence, one has to write \eqref{state1} as
\begin{equation}
 \psi_{\sf 1}(x,y,t)=\sum^{+\infty}_{l=-\infty}\psi_{l}(x,y)
 \sum^{+\infty}_{m=-\infty} J_{m} \left(\frac{u_j}{\tilde{\omega}}\right)\ e^{-iv_{F}(\epsilon+\tilde{\omega}(l+m))t}
\end{equation}
such that the eigenspinors $\psi_{l}(x,y)$ are solution of the following equation
\begin{equation}
H\psi_{l}(x,y)={{v_{F}}}(\epsilon+l\tilde{\omega})\psi_{l}(x,y)
\end{equation}
and the Hamiltonian is given by
\begin{equation}\label{eq 20}
H=v_{F}\left(%
\begin{array}{cc}
m^{+} & -i\frac{\sqrt{2}}{l_B} a^{-}\\
 i\frac{\sqrt{2}}{l_B} a^{+}  & m^{-} \\
\end{array}%
\right)
\end{equation}
where the shell operators for one mode 
\beq
a^{\pm}=\frac{l_B}{\sqrt{2}}\left(\mp\partial_{x}+k_y+\frac{x}{l_{B}^{2}}\right)
\eeq
are 
satisfying the canonical commutation relation $[a^{-}, a^{+}]=\mathbb{I}$, where we have defined 
the parameters $m^{\pm}= v\pm\tilde{\Delta}$
by rescaling the potential $v=V/v_{F}$ and energy gap $\tilde{\Delta}=\Delta/v_F$,
$l_B=\frac{1}{\sqrt{B_0}}$ is the magnetic length in the unit system $(\hbar=c=e=1)$.
We determine the eigenvalues and eigenspinors of the Hamiltonian \eqref{eq 20} by considering
the time-independent equation for the spinor
$\psi_{l}(x,y)=(\psi_{l}^{+}, \psi_{l}^{-})^{T}$ using the fact that the
transverse momentum $p_{y}$ is conserved to write $\psi_{l}(x,
y)=e^{ip_{y}y} \varphi_{l}(x)$ with $\varphi_{l}(x)=
(\varphi_{l}^+, \varphi_{l}^-)^{T}$. Then, the eigenvalue equation
\begin{equation}\label{eq 23}
H\left(%
\begin{array}{c}
  \varphi_{l}^+ \\
  \varphi_{l}^-\\
\end{array}%
\right)={{v_{F}}}(\epsilon+l\tilde{\omega})\left(%
\begin{array}{c}
  \varphi_{l}^+\\
  \varphi_{l}^-\\
\end{array}%
\right)
\end{equation}
gives the coupled equations
\begin{eqnarray}\label{eq 25}
  &&m^{+}\varphi_{l}^{+}-i\frac{\sqrt{2}}{l_{B}}a^{-}\varphi_{l}^-=(\epsilon+l\tilde{\omega})\varphi_{l}^+\\
&&\label{eq 26}
  i\frac{\sqrt{2}}{l_{B}}a^{+}\varphi_{l}^+ +
  m^{-}\varphi_{l}^{-}=(\epsilon+l\tilde{\omega})\varphi_{l}^-.
\end{eqnarray}
Inserting \eqref{eq 26} into \eqref{eq 25} we end up with  a
differential equation of second order for $\varphi_{l}^{+}$
\begin{equation}
(\epsilon+l\tilde{\omega}-m^{+})(\epsilon+l\tilde{\omega}-m^{-})\varphi_{l}^{+}=\frac{2}{l_{B}^{2}}a^-
a^{+}\varphi_{l}^{+}
\end{equation}
which is in fact the equation of a harmonic oscillator and
therefore we identify $\varphi_{l}^{+}$ with its eigenstates
$|n_{l}-1\rangle$ corresponding to the eigenvalues
\begin{equation}\label{eq34}
\tilde{\varepsilon}_{l}= 
\frac{1}{l_{B}}\sqrt{(\tilde{\Delta}
l_{B})^{2}+2n_{l}}
\end{equation}
where we have set
$\tilde{\varepsilon}_{l}=s'_{l}(\epsilon+l\tilde{\omega}-v)$,
$s'_{l}=\mbox{sign}(\epsilon+l\tilde{\omega}-v)$ correspond to
positive and negative energy solutions. The second spinor
component can  be derived from \eqref{eq 26} to obtain
\begin{equation}
\varphi_{l}^{-}=s'_{l}i\sqrt{\frac{\tilde{\varepsilon}_{l}
l_{B}-s'_{l} \tilde{\Delta} l_{B}}{\tilde{\varepsilon}_{l}
l_{B}+s'_{l} \tilde{\Delta} l_{B}}} \mid n_{l}\rangle.
\end{equation}
Introducing the parabolic cylinder functions $
D_{n_{l}}(x)=2^{-\frac{n_{l}}{2}}e^{-\frac{x^{2}}{4}}H_{n_{l}}\left(\frac{x}{\sqrt{2}}\right)$
we express the solution in region {\sf 1} as
\begin{eqnarray}
\psi_{\sf 1}(x,y,t) & =& e^{ik_{y}y}\sum^{+\infty}_{m,l=-\infty}\sum_{\pm}c^{\pm}_{l}\left(%
\begin{array}{c}
f_{l}
 D_{\left(\left(\tilde{\varepsilon}_{l} l_{B}\right)^{2}-(\tilde{\Delta} l_{B})^{2} \right)/2-1}
 \left(\pm \sqrt{2}\left(\frac{x}{l_{B}}+k_{y}l_{B}\right)\right) \\
  \pm i\tilde{f}_{l}
  D_{\left(\left(\tilde{\varepsilon}_{l} l_{B}\right)^{2}-\left(\tilde{\Delta} l_{B}\right)^{2}\right)/2}
 \left(\pm \sqrt{2}\left(\frac{x}{l_{B}}+k_{y}l_{B}\right)\right) \\
\end{array}%
\right) \nonumber \\
&&
J_{m-l}\left(\frac{u_j}{\tilde{\omega}}\right)e^{-iv_{F}(\epsilon+m\tilde{\omega})t }
\end{eqnarray}
where we have set
\begin{eqnarray}
&& f_{l}=\sqrt{\frac{\tilde{\varepsilon}_{l} l_{B}+s'_{l}
\tilde{\Delta} l_{B}}{\tilde{\varepsilon}_{l} l_{B}}}, \qquad
\tilde{f}_{l}=\frac{s'_{l}\sqrt{2}}{\sqrt{\tilde{\varepsilon}_{l}
l_{B}\left(\tilde{\varepsilon}_{l} l_{B}+s'_{l} \tilde{\Delta}
l_{B}\right)}}.
\end{eqnarray}
The above solutions will be used to compute some physical quantities in
our system such that the transmission probabilities and the associated GH shifts.

\section{GH shifts for  sidebands}

Note that for our system, as Dirac particles pass through a region
subject to time-harmonic potential, transitions from the
central band to sidebands (channels) at energies $\epsilon\pm
l\tilde{\omega}$ $(l = 0, 1, 2, \cdots)$ occur as particles
ex-change energy quanta with the oscillating field. Then to handle
wave propagation, we need to evaluate the transmission and reflection amplitudes,
which can be determined by matching different wave functions at
interfaces $0$ and $L$.  We write the continuity conditions 
\begin{eqnarray}\lb{sho0}
    \psi_{\sf 0}(0,y,t)=\psi_{\sf 1}(0,y,t) \lb{psix0}, \qquad
    \psi_{\sf 1}(L,y,t)=\psi_{\sf 2}(L,y,t) \lb{psixd}.
\end{eqnarray}
To simplify the notation, we use the following shorthand expressions
\begin{eqnarray}
&&\eta_{1,l}^{\pm}=D_{\frac{\left(\tilde{\varepsilon}_{l}
l_{B}\right)^{2}-\left(\tilde{\Delta} l_{B}\right)^{2}}{2}-1}
 \left(\pm \sqrt{2}k_{y}l_{B}\right)\lb{sho1}\\
&& \xi_{1,l}^{\pm}= D_{\frac{\left(\tilde{\varepsilon}_{l}
l_{B}\right)^{2}-\left(\tilde{\Delta} l_{B}\right)^{2}}{2}}
  \left(\pm \sqrt{2}k_{y}l_{B}\right)\lb{sho2}\\
&& \eta_{2,l}^{\pm}=D_{\frac{\left(\tilde{\varepsilon}_{l}
l_{B}\right)^{2}-\left(\tilde{\Delta} l_{B}\right)^{2}}{2}-1}
 \left[\pm \sqrt{2}\left(\frac{L}{l_{B}}+k_{y}l_{B}\right)\right]\lb{sho3}\\
&& \xi_{2,l}^{\pm}= D_{\frac{\left(\tilde{\varepsilon}_{l}
l_{B}\right)^{2}-\left(\tilde{\Delta} l_{B}\right)^{2}}{2}}
  \left[\pm \sqrt{2}\left(\frac{L}{l_{B}}+k_{y}l_{B}\right)\right].\lb{sho4}
\end{eqnarray}
To derive different physical quantities, one can explicitly write
(\ref{sho0}) making use the fact that the basis
$\{e^{imv_{F}\tilde{\omega} t}\}$ is orthogonal. Thus at the interface
$x=0$, one finds
\begin{eqnarray}
&& \delta_{m0}+r_{m}=\sum^{+\infty}_{l=-\infty}
\left(c^{+}_{l}f_l\eta_{1,l}^{+}
+c^{-}_{l}f_l\eta_{1,l}^{-}\right)
J_{m-l}\left(\frac{u_j}{\tilde{\omega}}\right) \lb{eqx01}\\
&& \delta_{m0}\alpha_{m}-r_{m}\frac{1}{\alpha_{m}}=
\sum^{+\infty}_{l=-\infty}
\left(c^{+}_{l}i\tilde{f}_{l}\xi_{1,l}^{+}-c^{-}_{l}i\tilde{f}_{l}\xi_{1,l}^{-}\right)
J_{m-l}\left(\frac{u_j}{\tilde{\omega}}\right) \lb{eqx02}
\end{eqnarray}
and at $x=L$, we have 
\begin{eqnarray}
&& t_{m}e^{ik^{'}_{m}d}+
b_{m}e^{-ik^{'}_{m}d}=\sum^{+\infty}_{l=-\infty} \left(
 c^{+}_{l}f_l\eta_{2,l}^{+} +c^{-}_{l}f_l\eta_{2,l}^{-}\right) J_{m-l}\left(\frac{u_j}{\tilde{\omega}}\right) \lb{eqxd1} \\
&&t_{m}\beta_{m}e^{ik^{'}_{m}d}-b_{m}\frac{1}{\beta_{m}}e^{-ik^{'}_{m}d}
= \sum^{+\infty}_{l=-\infty}
\left(c^{+}_{l}i\tilde{f}_{l}\xi_{2,l}^{+}
-c^{-}_{l}i\tilde{f}_{l}
\xi_{2,l}^{-}\right)J_{m-l}\left(\frac{u_j}{\tilde{\omega}}\right). \lb{eqxd2}
\end{eqnarray}
It is convenient to write (\ref{eqx01}-\ref{eqxd2}) in matrix form
\begin{eqnarray}
\left(%
\begin{array}{c}
  \Xi_{0} \\
  \Xi_{0}^{'} \\
\end{array}%
\right)=\left(%
\begin{array}{cc}
 { \mathbb M11} &{\mathbb M12} \\
 {\mathbb M21} &{ \mathbb M22} \\
\end{array}%
\right)\left(%
\begin{array}{c}
  \Xi_{2} \\
  \Xi_{2}^{'}\\
\end{array}%
\right)={\mathbb M}\left(%
\begin{array}{c}
  \Xi_{2} \\
 \Xi_{2}^{'} \\
\end{array}%
\right)
\end{eqnarray}
such that 
${\mathbb M}={\mathbb
M(0,1)}\cdot{\mathbb M(1,2)}$ and ${\mathbb M(j,j+1)}$ are
transfer matrices that couple the wave function in the $j$-th
region to the wave function in the $(j + 1)$-th region, with 
\begin{eqnarray}
&&{\mathbb M(0,1)}=\left(%
\begin{array}{cc}
  {\mathbb I}& {\mathbb I} \\
{\mathbb N_{1}^{+}} &{\mathbb N_{1}^{-}} \\
\end{array}%
\right)^{-1}
\left(%
\begin{array}{cc}
  {\mathbb C_{1}^{+}} & {\mathbb C_{1}^{-}} \\
 {\mathbb G_{1}^{+}} & {\mathbb G_{1}^{-}} \\
\end{array}%
\right)\\
&&{\mathbb M(1,2)}=\left(%
\begin{array}{cc}
  {\mathbb C_{2}^{+}} & {\mathbb C_{2}^{-}} \\
  {\mathbb G_{2}^{+}} & {\mathbb G_{2}^{-}} \\
\end{array}%
\right)^{-1}
\left(%
\begin{array}{cc}
  {\mathbb I}& {\mathbb I} \\
{\mathbb N_{2}^{+}} &{\mathbb N_{2}^{-}} \\
\end{array}%
\right)\left(%
\begin{array}{cc}
  {\mathbb K^{+}}&{\mathbb O}  \\
{\mathbb O} &{\mathbb K^{-}} \\
\end{array}%
\right)
\end{eqnarray}
where we have defined the quantities
\begin{eqnarray}
&&\left({\mathbb
N_{1}^{\pm}}\right)_{m,l}=\pm\left(\alpha_{m}\right)^{\pm
1}\delta_{ml} \lb{eqn1}
\\
&&\left({\mathbb
C_{\tau}^{\pm}}\right)_{m,l}=f_{l}\eta_{\tau,l}^{\pm}J_{m-l}\left(\frac{u_j}{\tilde{\omega}}\right)
\\
&&\left({\mathbb G_{\tau}^{\pm}}\right)_{m,l}=\pm
i\tilde{f}_{l}\xi_{\tau,l}^{\pm}J_{m-l}\left(\frac{u_j}{\tilde{\omega}}\right)
\\
&&\left({\mathbb  K^{\pm}}\right)_{m,l}=\pm e^{\pm
idk_{m}^{'}}\delta_{ml}
\\
&&\left({\mathbb
 N_{2}^{\pm}}\right)_{m,l}=\pm\left(\beta_{m}\right)^{\pm
1}\delta_{ml}\lb{eqn2}
\end{eqnarray}
with the null matrix denoted by ${\mathbb O}$ and  ${\mathbb
I}$ is the unit matrix. We assume a particle propagating from
left to right with 
energy $\epsilon$ then $\tau=(1,2)$, $\Xi_{0}=\{\delta_{0l}\}$ and
$\Xi_{2}^{'}=\{b_{m}\}$ is the null vector, whereas the vectors
for wave transmission and reflection are $\Xi_{2}=\{t_{l}\}$, and
$\Xi_{0}^{'}=\{r_{l}\}$, respectively. From the above considerations, one can
easily obtain the relation $\Xi_{2}=\left({ \mathbb
M11}\right)^{-1} \cdot\Xi_{0}$. The minimum number $N$ of
sidebands that need to be considered is determined by the strength
of the oscillation frequency, $N>\frac{u_j}{\tilde\omega}$, and the infinite series for $T$
can then be truncated to consider a finite number of terms starting
from $-N$ up to $N$. Furthermore, analytical results are obtained
if we take small values of {$\frac{u_j}{\tilde\omega}$}
and include only the first two sidebands at energies $\epsilon\pm
\tilde{\omega}$ along with the
central band at energy $\epsilon$, such as
\begin{equation}
t_{-N+k}={ \mathbb M^{'}}\left[k+1, N+1\right]\lb{eqt}
\end{equation}
where $k=0, 1, 2, \cdots, 2N$ and ${ \mathbb M^{'}}$ is a matrix
element of $\left({ \mathbb M11}\right)^{-1}$.

In the forthcoming analysis we truncate (\ref{eqt}) retaining only the terms corresponding to the central
and first two sidebands, namely $l=0,\pm 1$.  This is justified for the driving amplitudes we consider because they
are weak enough so that this approximation does not break. We can proceed as before to derive transmission amplitudes
\begin{equation}
  t_{-1}={ \mathbb M^{'}}[1,2], \qquad t_{0}={ \mathbb M^{'}}[2,2], \qquad t_{1}={ \mathbb M^{'}}[3,2].
\end{equation}
Now for a null amplitude of oscillating potential
 ($U_j=v_F u_j=0$), we get  only the
transmission amplitude $t_{0}$ for the central band. It can be
calculated as 
\begin{equation}
t_{0}=\rho_{0}e^{i\varphi^{t}_{0}}=\frac{i2f_{0}\tilde{f}_{0}\cos\phi_{0}}{e^{idk^{'}_{0}}\left(\chi_{0}-if_{0}\tilde{f}_{0}\Omega_{0}\right)}
\left(\eta^{+}_{2,0}\xi^{-}_{2,0}+\eta^{-}_{2,0}\xi^{+}_{2,0}\right)
\end{equation}
where the phase shift
$\varphi^{t}_{0}=\arctan\left(i\frac{t^{\ast}_{0}-t_{0}}{t^{\ast}_{0}+t_{0}}\right)$,
$\rho_{0}=\sqrt{\mbox{Im}[t]^{2}+\mbox{Re}[t]^{2}}$ and different
quantities are
\begin{eqnarray}
&&
\chi_{0}=f^{2}_{0}\left(\eta^{+}_{1,0}\eta^{-}_{2,0}-\eta^{-}_{1,0}\eta^{+}_{2,0}\right)e^{i\left(\theta_{0}-\phi_{0}\right)}-
\tilde{f}^{2}_{0}\left(\xi^{+}_{1,0}\xi^{-}_{2,0}-\xi^{-}_{1,0}\xi^{+}_{2,0}\right)
\\
&&
\Omega_{0}=\left(\left(\eta^{+}_{2,0}\xi^{-}_{1,0}+\eta^{-}_{2,0}\xi^{+}_{1,0}\right)e^{i\theta_{0}}
+\left(\eta^{+}_{1,0}\xi^{-}_{2,0}+\eta^{-}_{1,0}\xi^{+}_{2,0}\right)e^{-i\phi_{0}}\right).
\end{eqnarray}
As a result, the transmission  probabilities are finally expressed
as 
\beq
T_{l}=\frac{k^{'}_{l}}{k_{0}}\rho_{l}^{2}, \qquad  l=0,\pm 1.
\eeq

The Goos-H\"anchen (GH) shifts in graphene can be analyzed by
considering the  incident, reflected and  transmitted beams around
some transverse wave vector $k_y = k_{y_0}$ together with the  angle of
incidence $\phi_{l}(k_{y_{0}})\in [0, \frac{\pi}{2}]$, denoted by
the subscript $0$. These can be expressed in integral forms 
for incident 
\begin{eqnarray}
   \Psi_{in}(x,y) &=& \int_{-\infty}^{+\infty}dk_y\ f(k_y-k_{y_0})\ e^{i(k_{0}(k_y)x+k_yy)}
            \begin{pmatrix}
              {1} \\
              {e^{i\phi_{0}(k_{y})}}
            \end{pmatrix}
          \label{eq 79}
\end{eqnarray}
and  reflected beams 
\begin{eqnarray}
\Psi_{re}(x,y) &=& \int_{-\infty}^{+\infty}dk_y\ r_{l}(k_y)\
f(k_y-k_{y_0})\ e^{i(-k_{l}(k_y)x+k_yy)}
            \begin{pmatrix}
              {1} \\
              {-e^{-i\phi_{l}(k_{y})}} \\
            \end{pmatrix}
          \label{refl}
\end{eqnarray}
where the reflection amplitude is defined by
$r_{l}(k_y)=|r_{l}|e^{i\varphi_{l}^{r}}$. This result is reached
by writing the $x$-component of wave vector $k_l$ as well as
$\phi_l$ in terms of $k_y$ such that each spinor plane wave is a
solution of \eqref{ham1} and $f(k_y - k_{y_0})$ is the angular spectral
distribution. We can approximate the $k_y$-dependent terms by a
Taylor expansion around $k_y$, retaining only the first order term
to get
\begin{eqnarray}
&&\phi_{l}(k_{y})\approx
\phi_{l}(k_{y_{0}})+\frac{\partial\phi_{l}}{\partial
k_{y}}\Big|_{k_{y_{0}}}(k_{y}-k_{y_{0}})
\\
&& k_{l}(k_{y})\approx k_{l}(k_{y_{0}})+\frac{\partial
k_{l}}{\partial k_{y}}\Big|_{k_{y_{0}}}(k_{y}-k_{y_{0}}).
\end{eqnarray}
Finally, the transmitted beams are
\begin{eqnarray}
\Psi_{tr}(x,y) &=& \int_{-\infty}^{+\infty}dk_y\ t_{l}(k_y)\
f(k_y-k_{y_0})\ e^{i(k^{'}_{l}(k_y)x+k_yy)}
            \begin{pmatrix}
              {1} \\
              {e^{i\theta_{l}(k_{y})}} \\
            \end{pmatrix}
          \label{trans}
\end{eqnarray}
where the transmission coefficient is
$t_{l}(k_y)=|t_{l}|e^{i\varphi_{l}^{t}}$.

 The stationary-phase approximation indicates that the GH shifts are equal to the negative gradient
of transmission phase with respect to $k_y$. To calculate the GH
shifts of the transmitted beams through our system, according to
the stationary phase method \cite{Bohm}, we adopt the definition
\cite{xChen1, mMekkaoui}
\begin{equation}
        S_{l}^{t}=- \frac{\partial \varphi_{l}^{t}}{\partial
        k_{y}}\Big|_{k_{y_0}}, \qquad S_{l}^{r}=- \frac{\partial \varphi_{l}^{r}}{\partial
        k_{y}}\Big|_{k_{y_0}}.\label{eq 51}
 \end{equation}
Assuming a finite-width beam with a Gaussian shape,
$f(k_y-k_{y_0})=w_y\exp[-w_{y}^2(k_y-k_{y_0})^2]$ around $k_{y_0}$,
where $w_{y}=w\sec\phi_l$,  and $w$ is the half beam width at
waist. We can now evaluate the Gaussian integral to obtain the spatial
profile of the incident beam, by expanding $\phi_l$ and $k_{l}$ to
first order around $k_{y_0}$ while satisfying the condition
$\delta\phi_l=\lambda_{F}/(\pi w)\ll 1$
where $\lambda_{F}$ is the Fermi wavelength. Comparison of the
incident and transmitted beams suggests that the displacements
$\sigma_{\pm}$ of up and down spinor components are both equal to
$\partial \varphi_{l}^{t}/\partial k_{y_0}$ and the average
displacement is
\begin{equation}
S_{l}^{t}=\frac{1}{2}(\sigma^{+}+\sigma^{-})=- \frac{\partial
\varphi_{l}^{t}}{\partial
        k_{y}}\Big|_{k_{y_{0}}}.
\end{equation}
It should be noted that when the above-mentioned condition is
satisfied, that is, the stationary phase method is valid
\cite{xChen1}, the definition \eqref{eq 51} can be applied to
any finite-width beam, the shape does not necessarily need to be Gaussian-shaped.

\section{Results and Discussions}

\begin{figure}[ht]
        \centering
        \subfloat[]{
            \centering
            \includegraphics[height=4.0cm] 
            {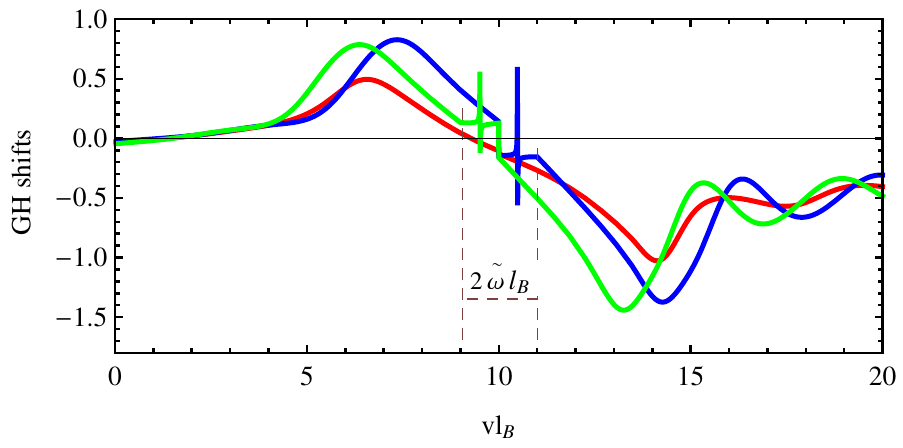} 
            \label{db.3}
        }\subfloat[]{
            \centering
            \includegraphics[height=4.0cm] 
            {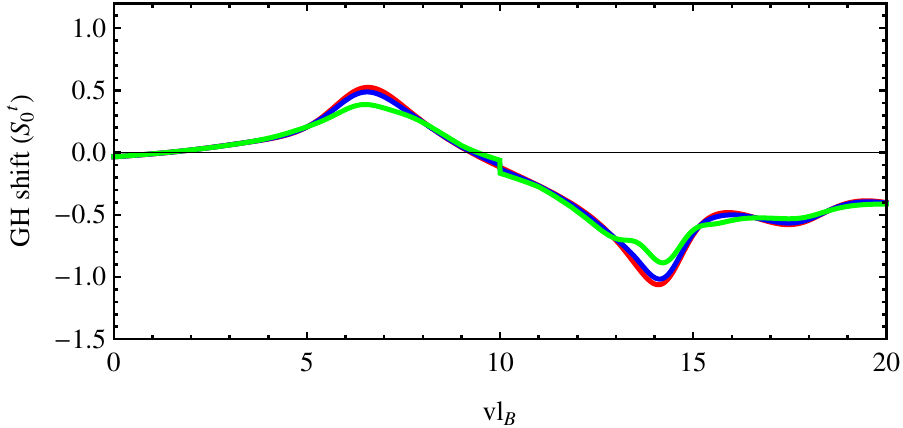}
            \label{db.4}}
    \caption{\sf{(color online) 
    The GH shifts in transmission  as
a function of $vl_B$ with  $k_{y} l_{B}=2$, $\frac{L}{l_{B}}=0.8$, $\tilde{\omega}
l_{B}=1$ $\tilde{\Delta} l_{B}=0$ and $\epsilon l_{B}=10$. {\color{blue}(a)}: Central band $S_{0}^{t}$ in red,
the first two sidebands $S_{1}^{t}$ in blue and $S_{-1}^{t}$ in green
 with $\alpha=0.4$. {\color{blue}(b)}: Central band $S_{0}^{t}$. For different values of $\frac{u_j}{\tilde\omega}$: 0 (red line),
$0.45$ (blue line), $0.85$ (green line).}}
    \label{fig2}
\end{figure}

In this section, we study the effect of the magnetic field on the
Goos-H\"anchen (GH) shifts for Dirac fermions in gaped graphene  that is
subject to an additional time periodic oscillating potential in the region of space with the potential barrier. We numerically
evaluate the GH shifts in transmission for the central band
$S_{0}^{t}$  and two first sidebands $S_{\pm 1}^{t}$ as a function
of the parameters of the graphene single barrier
oscillation frequency $\tilde{\omega} l_B$ and amplitude
$\frac{u_j}{\tilde\omega}$, in addition to the energy
$\epsilon l_B$, the $y$-component of the wave vector $k_y l_B$,
the energy gap $\tilde{\Delta} l_B$ and the strength of static
potential barrier $v l_B$. 
To understand these effects, let us
consider Figure \text{\ref{fig2}} where we
study the GH shifts in transmission versus the potential
barrier $v l_B$ in the gapless graphene
region where $\tilde{\Delta} l_{B}=0$, the frequency
$\tilde{\omega} l_{B}=1$, and the energy $\epsilon l_{B}=10$, $k_{y} l_{B}=2$,
$\frac{L}{l_{B}}=0.8$. Figure \text{\ref{fig2}}{\color{blue}a} shows the
GH shifts in transmission for the central band $S_{0}^{t}$ (red line) and
first two sidebands $S_{-1}^{t}$ (green line), $S_{1}^{t}$ (blue
line) where the value $\frac{u_j}{\tilde\omega}=0.4$,
the GH shifts in transmission changes sign at the Dirac points
$vl_B=\epsilon l_B+l\tilde{\omega}$ ($l=0$ for band central and
$l=\pm 1$ for first two sidebands). It is clearly seen that they are
strongly dependent on the barrier heights and frequency
$\tilde{\omega} l_B$. Figure \text{\ref{fig2}}{\color{blue}b}
shows the GH shifts in transmission for the central band for
different values of $\frac{u_j}{\tilde\omega}=\{0, 0.45, 0.85\}$. One can notice that,
at the Dirac points $vl_{B}=\epsilon l_{B}$, the GH shifts change
their sign. We observe that the GH shifts for central band
$S_{0}^{t}$ in the oscillating barrier decreases if $\frac{u_j}{\tilde\omega}$
increases. We also notice that the GH shifts can have either sign, positive and negative
in Figures \text{\ref{fig2}}{\color{blue}a} and \text{\ref{fig2}}{\color{blue}b}.

\begin{figure}[ht]
        \centering
        \subfloat[$\tilde{\Delta}=0$]{
            \centering
            \includegraphics[height=4.0cm]{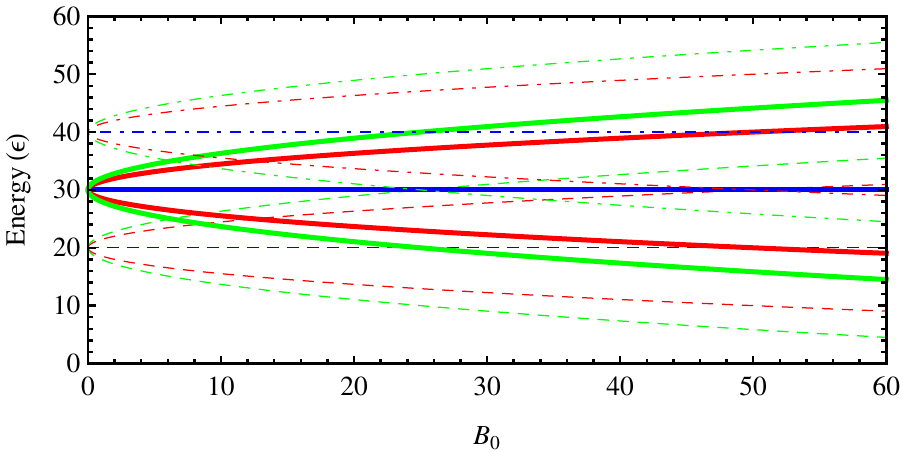}
            \label{db.3}
        }\subfloat[$\tilde{\Delta}=10$]{
            \centering
            \includegraphics[height=4.0cm]{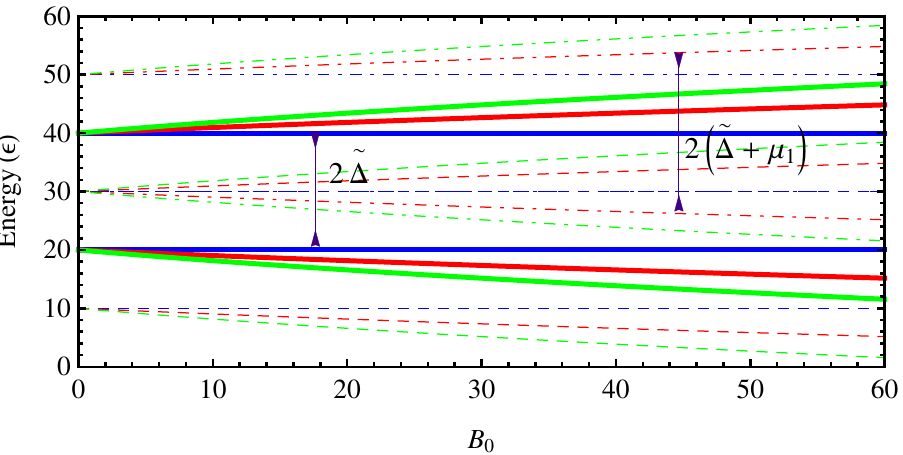}
            \label{db.4}}
    \caption{\sf{(color online) 
    Graphs depicting the energy $\epsilon$ as a function of magnetic field $B$ with the energy gap {\color{blue}(a)}: $\tilde{\Delta}=0$ and {\color{blue}(a)}:  $\tilde{\Delta}=10$. For
    the frequency $\tilde{\omega}=10$,
     potential $v=30$, quantum
number $n=0$ (blue line), $n=1$ (red line), $n=2$  (green line)  and the  modes
$l=0$ (color thick), $l=1$ (color dashed), $l=-1$ (color dot-dashed).}}
    \label{fig3}
\end{figure}

Now let us investigate what happens if we introduce a gap in
the intermediate region $0\leq x\leq L$,  which is also subjected to a magnetic
field. Note that, the gap is introduced as shown in Figure \text{\ref{fig1}} and
therefore it affects the system energy according to the solution
of the energy spectrum obtained in region $1$. From \eqref{eq34}, we
obtain the energy modulation due the oscillating potential as
shown in Figure \text{\ref{fig3}} as function of the magnetic field $B$ with
$v=30$, $w=10$, $l=$\{-1: (\text{color dot-dashed}), 0: (\text{color thick}), 1:
(\text{color dashed})\}
and $n=$\{0: (\text{blue line}), 1: (\text{red line}), 2: (\text{green line})\}
for gapless  $\tilde{\Delta}=0$
in Figure \text{\ref{fig3}}{\color{blue}a} and gap
$\tilde{\Delta}=10$ in Figure \text{\ref{fig3}}{\color{blue}b}. It is clearly seen that the difference of energy is
$\epsilon(n+1; l)-\epsilon(n; l)=\tilde{\omega}$, which is independent of the quantum
number $n$. For $n=0$, we have just a modulation
of the energy with different quantum number $l\tilde{\omega}$ with
$l=0,\pm 1$. However for $n=1,2$ the energy behavior is completely
changed and for each $l$ value the energy is split into two
values, which can be seen like a left of degeneracy of levels. It
is clearly seen that absorbing energy quantum $\tilde{\omega}$
produces inter-level transitions. Because of Pauli principle a
particle with energy $\epsilon$ can absorb an energy quantum
$\tilde{\omega}$ if and only if the state with energy
$\epsilon+\tilde{\omega}$ is empty. We observe that in Figure \text{\ref{fig3}}{\color{blue}b}
the difference of energy $\epsilon^{+}$ and $\epsilon^{-}$ for
quantum number $n=0$ and the value $l=0,\pm 1$ is
$\epsilon^{+}(n=0; l=0, \pm 1)-\epsilon^{-}(n=0; l=0, \pm
1)=2\tilde{\Delta}$. The difference of energy $\epsilon^{+}$ and
$\epsilon^{-}$ for quantum number $n=1, 2$ and the value $l=0,\pm
1$ is $\epsilon^{+}(n=1,2; l=0, \pm 1)-\epsilon^{-}(n=1, 2; l=0,
\pm 1)=2(\tilde{\Delta}+\mu_{n})$ with
$\mu_{n}=\epsilon^{+}(n=1,2; l=0, \pm 1)-\epsilon^{+}(n=0; l=0,
\pm 1)$.

\begin{figure}[ht]
        \centering
        \subfloat[]{
            \centering
            \includegraphics[height=4.0cm]{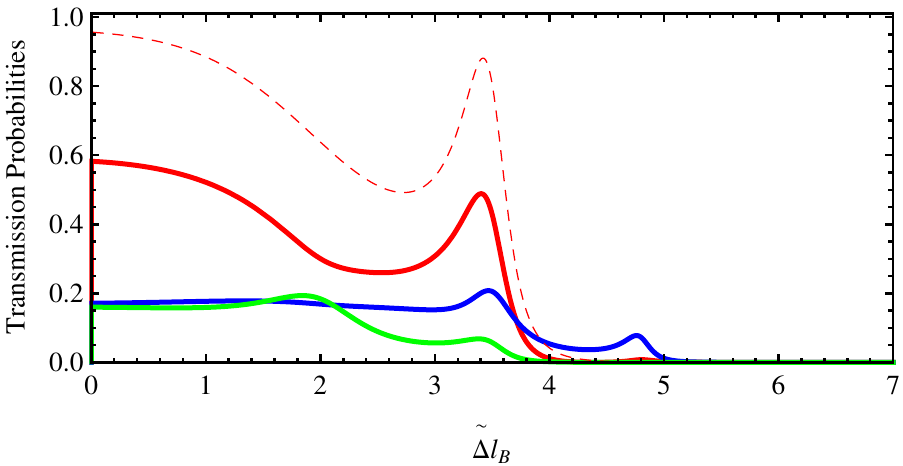}
            \label{db.3}
        }\subfloat[]{
            \centering
            \includegraphics[height=4.0cm]{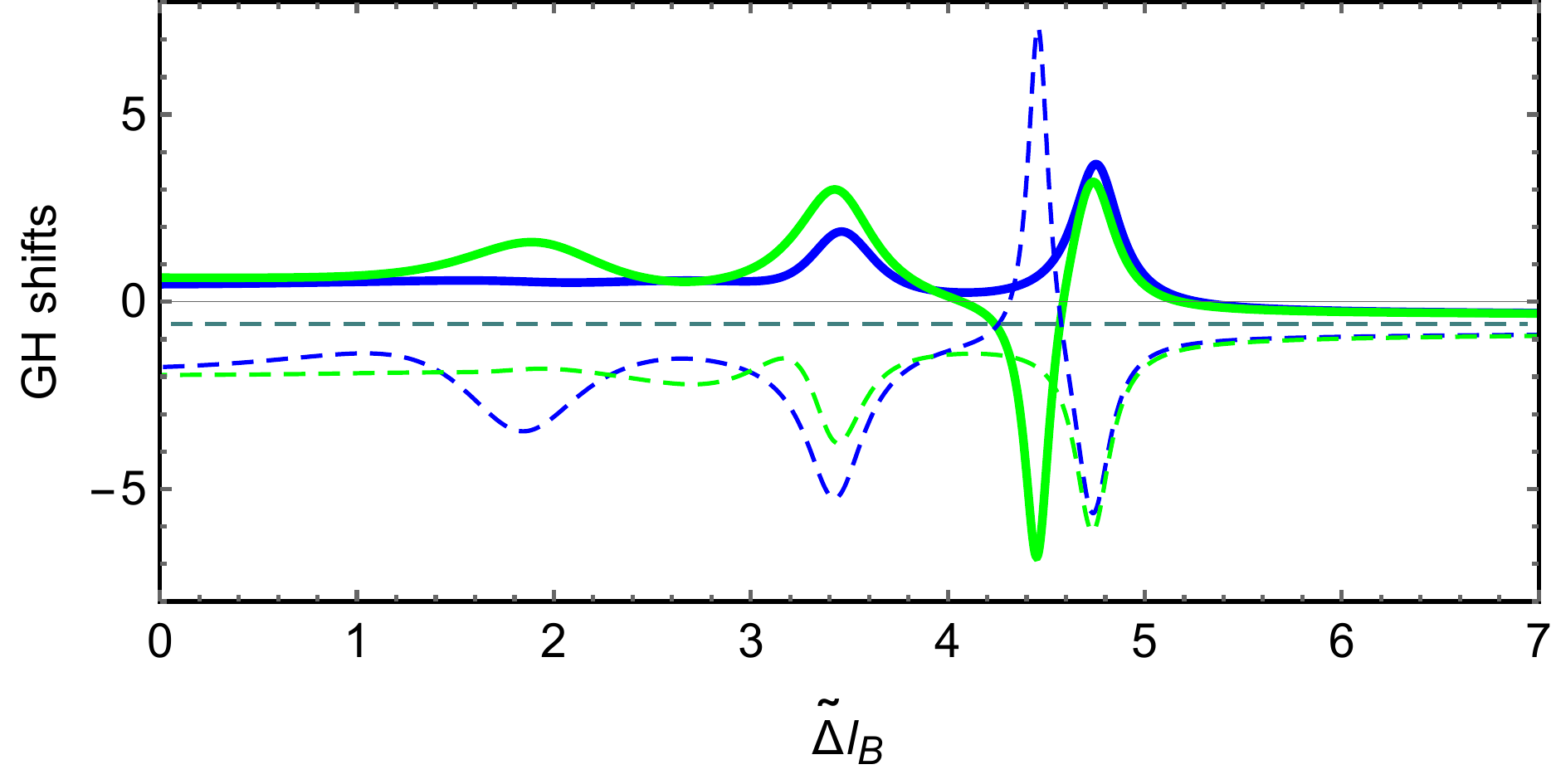}
            \label{db.4}}
    \caption{\sf{(color online) {\color{blue}(a)}:
 The transmission
probabilities $T_{l}$ for central band and first tow sidebands and {\color{blue}(b)}: GH
shifts in transmission $S_{\pm 1}^t$ for first tow sidebands 
as a function of energy gap $\tilde{\Delta} l_{B}$ with
$\frac{u_j}{\tilde\omega}=0$ (not oscillating barrier), $\frac{u_j}{\tilde\omega}=0.65$ (oscillating barrier), $k_{y} l_{B}=2$,
$\frac{L}{l_{B}}=1.4$, $\tilde{\omega} l_{B}=1$ and ($\{\epsilon
l_{B}=12; vl_{B}=7\})/(\{\epsilon
l_{B}=12; vl_{B}=7\}, \{\epsilon
l_{B}=7; vl_{B}=12\}$). For no oscillating barrier  $T_{0}$ (red dashed) and for oscillating
barrier $T_{0}$ (red line), $T_{-1}$
(green line), $T_{1}$ (blue line). }}
    \label{fig4}
\end{figure}

In Figure \text{\ref{fig4}} we present the transmission probabilities  $T_{l}$ 
for the central band and first two
sidebands together with 
the GH shifts in transmission $S_{\pm 1}^t$ for first two
sidebands 
as a function of the energy gap $\tilde{\Delta}l_B$
for $\frac{u_j}{\tilde\omega}=0.65$ (oscillating magnetic barrier) along with
the results for a static barrier and
specific values $k_{y} l_{B}=2$, $\frac{L}{l_{B}}=1.4$,
$\tilde{\omega} l_{B}=1$. In Figure \text{\ref{fig4}}{\color{blue}a} for  $\frac{u_j}{\tilde\omega}=0,65$, we observe that  
the maximum value of $T_{0}$
decreases at the expense of transmission sidebands $T_{\pm 1}$. We
notice that the sum of the three transmissions $T_{0,-1,1}$
converges whenever towards transmission $T_{0}$ for $\frac{u_j}{\tilde\omega}=0$. In fact,
under the condition $\tilde{\Delta}l_B> |\epsilon l_B+l\tilde{\omega}l_B-v l_B|$ every incoming state is fully
reflected. In Figure \text{\ref{fig4}}{\color{blue}b}, the GH shifts $S_{\pm 1}^t$  in transmission for first two sidebands
in the propagating case can be enhanced by opening a gap at the Dirac point. This  computation has been performed keeping the
parameters $\frac{u_j}{\tilde\omega}=0.65$, $k_{y} l_{B}=2$, $\frac{L}{l_{B}}=1.4$,
$\tilde{\omega} l_{B}=1$ and making different choices for the
energy $\epsilon l_B$ and potential $vl_B$. For the configuration
$\{\epsilon l_B=12, vl_B=7\}$, we can still have positive shifts
while for configuration $\{\epsilon l_B=7, vl_B=12\}$ the GH
shifts are negative. 
The GH shifts in
transmission for the first  two sidebands $S_{1}^t$ (blue line) and $S_{-1}^t$ (green line) did not
vanish and decrease with increasing $\tilde{\Delta}l_B$ for
$s=\text{sign}(\epsilon l_B-vl_B)=-1$ as well as increases with increasing
$\tilde{\Delta}l_B$  for $s=\text{sign}(\epsilon l_B-vl_B)=1$. It is clearly seen that the GH shifts can be enhanced by a
certain gap opening. Indeed, by increasing the gap we observe that the gap of transmission becomes
broader, changing the transmission resonances and the modulation of the GH shifts. Note that for a
certain energy gap $\tilde{\Delta}l_B$, there is total reflection and therefore the GH shifts in transmission $S_{\pm 1}^t$ do not
vanish.

\begin{figure}[ht]
        \centering
        \subfloat[]{
            \centering
            \includegraphics[height=7cm]{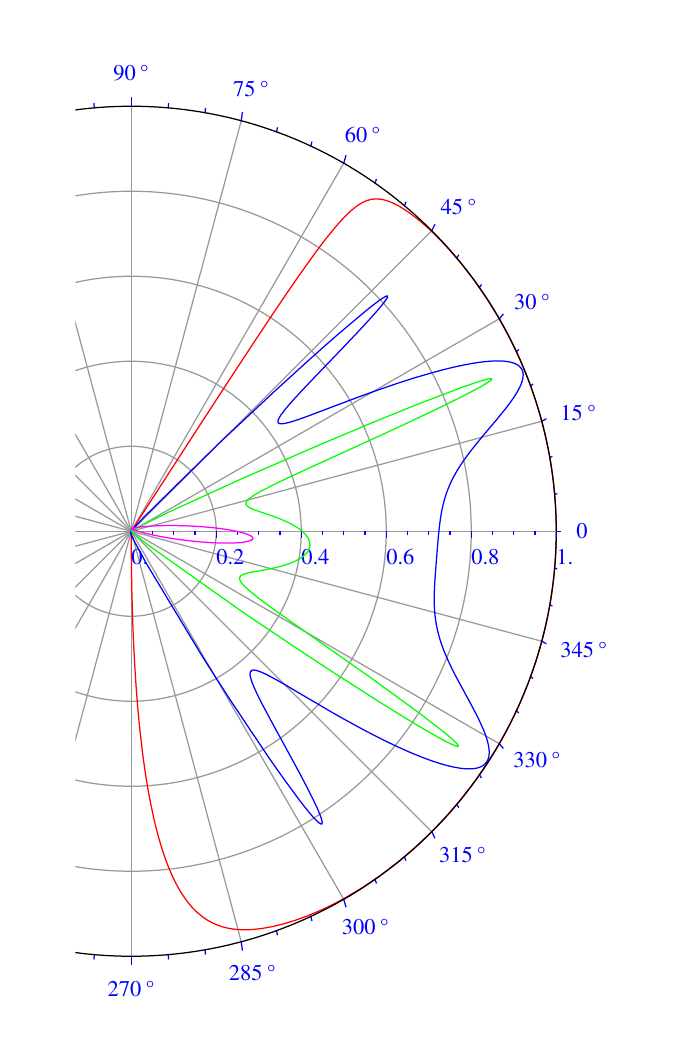}
            \label{fig3a}\ \ \ \ \
        }\subfloat[]{
            \centering
            \includegraphics[height=7cm]{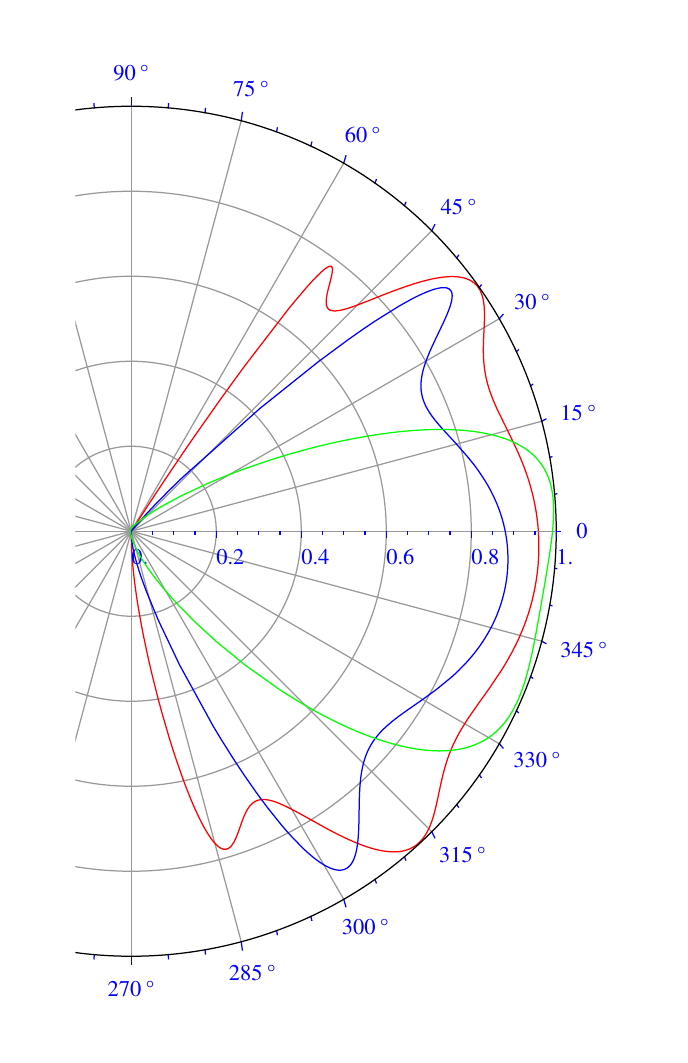}
            \label{fig3b}}
    \caption{\sf{(color online) Polar graphs depicting the transmission
probability $T_{0}(\phi_{0})$ for a magnetic barrier of $\frac{u_j}{\tilde\om}=0$ with  $\frac{L}{l_{B}}=1.2$, $vl_{B}=0$.
{\color{blue}{(a)}}\color{black}{:} For various gap
$\tilde{\Delta} l_{B}=\{0, 4, 6, 7.1\}$ and $\epsilon l_{B}=7.5$.
{\color{blue}{(b)}}\color{black}{:} For various energy $\epsilon l_{B}=\{1.8, 3.5, 5, 7.5\}$
 and $\tilde{\Delta} l_{B}=2$. }}
    \label{fig5}
\end{figure}

In Figure \text{\ref{fig5}}, we plot  the transmission probabilities in central band $T_{0}$ for magnetic barrier $\frac{u_j}{\tilde\om}=0$
as a function of the incidence angle $\phi_{0}$ for specific values of $\frac{L}{l_{B}}=1.2$, $vl_{B}=0$. We show
that it is possible to confine massless Dirac fermions in  graphene sheet by inhomogeneous
magnetic field and the induced gap $\tilde{\Delta}$. The
outermost circle corresponds to full transmission total $T_{0}=1$, while
the origin of this plot represents zero transmission $T_{0}=0$. In Figure \text{\ref{fig5}}{\color{blue}a} we show how the transmission is
affected by the effective mass term reflected by $\tilde{\Delta}l_B=\{0, 4, 6, 7.1\}$ and the
parameters $\epsilon l_{B}=7.5$, the transmission decreases sharply as we increase the energy gap $\tilde{\Delta}l_B$,
there is no transmission possible. We notice that
for certain incidence angles the transmission is not allowed, in
fact for $\epsilon l_B\leq L/l_B-l\tilde{\omega}l_B$ all waves are completely
reflected. It is worth mentioning that the transmission is
uniquely defined by the incidence angle $\phi_{0}$. Each radial line
represents a given incidence angle and intersects the transmission
curve at one point. In Figure \text{\ref{fig5}}{\color{blue}b} for the energy gap $\tilde{\Delta} l_{B}=2$
and various energy $\epsilon l_{B}=\{1.8, 3.5, 5, 7.5\}$, we
see that the transmission
vanishes for $\epsilon l_B\leq 1.2$.

It is well known that graphene has a zero band gap because the
Dirac-Weyl Hamiltonian, that models graphene, describes massless quasiparticles \cite{Novoselov, Zhang}
and hence allows for Klein tunneling. However, electronic components such as electronic switches,
diodes and transistors, require that the current can be
cut off/on. It is necessary then to induce an energy gap in
graphene in order to control the current flow. Therefore, 
we also investigated the influence of the energy gap on the GH shifts
in 
Figure \text{\ref{fig6}} where 
 $S_{l}^{t}$ for the central band and first two
sidebands 
is plotted as a function of the potential $\epsilon l_B$
for $\frac{u_j}{\tilde\om}=0.4$ and
specific values $k_{y}l_B=2$, $\epsilon l_B=10$,
$d/l_B = 1.5$, $\tilde{\omega}l_B=1$, $\tilde{\Delta}l_B=1$,  different
energy gaps such that $\tilde{\Delta}l_B=1$  in Figure
\text{\ref{fig6}}{\color{blue}a} and $\tilde{\Delta}l_B=3$  in Figure \text{\ref{fig6}}{\color{blue}b}.
From these, we see that the region of
weak GH shifts becomes wider with increased energy gap. Then, we can control the positive and negative
GH shifts by changing the $y$-directional wave vector $k_y/l_B$ or the energy gap $\tilde{\Delta} l_{B}$. In other words,
we can control the directions of
the carriers at the interface of the graphene barrier by adjusting $k_y/l_B$ or $\tilde{\Delta} l_{B}$.
The GH shifts still change sign and the absolute value of the maximum of the shifts increased as well. It is clearly seen that 
$S_{l}^t$
are oscillating between negative and positive values around
the critical point $\epsilon l_B=vl_B-l\tilde{\omega}l_B$ $(l=0,\pm 1)$.\\

\begin{figure}[h!]
\centering
\includegraphics [height=4cm]{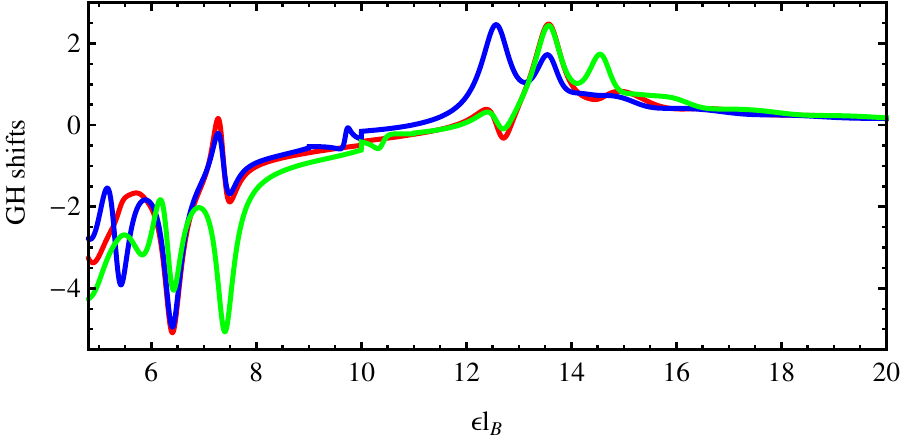}
\includegraphics[height=4cm]{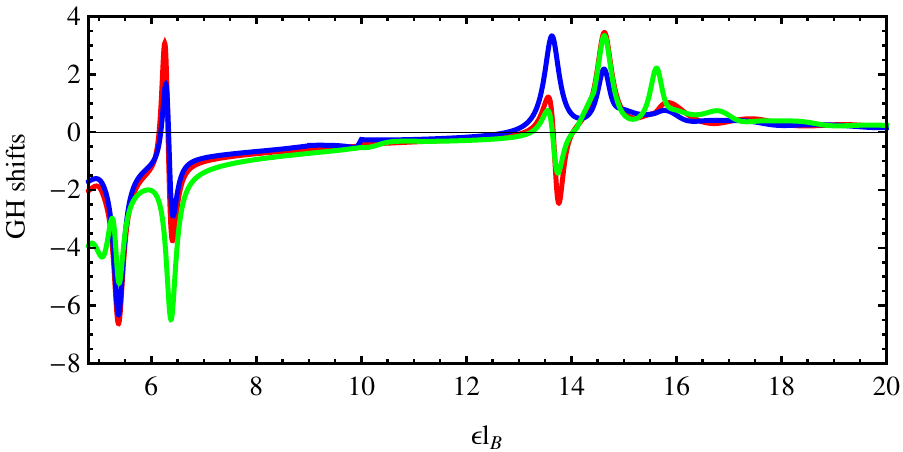}
 \caption{\sf{(color online) The GH shifts in transmission for central band and first
tow sidebands   as a function of the energy
$\epsilon l_B$ for $\frac{u_j}{\tilde\om}=0.4$, 
 $k_{y} l_{B}=2$, $\epsilon l_{B}=10$, $\frac{L}{l_{B}}=1.5$, $\tilde{\omega} l_{B}=1$
 with {\color{blue}(a)}: $\tilde{\Delta}l_{B}=1$ and {\color{blue}(b)}: $\tilde{\Delta}l_{B}=3$. 
$S_{0}^{t}$ (red line), $S_{-1}^{t}$
(green line), $S_{1}^{t}$ (blue line).}}\lb{fig6}
\end{figure}

\begin{figure}[ht]
        \centering
        \subfloat[]{
            \centering
            \includegraphics[height=4cm]{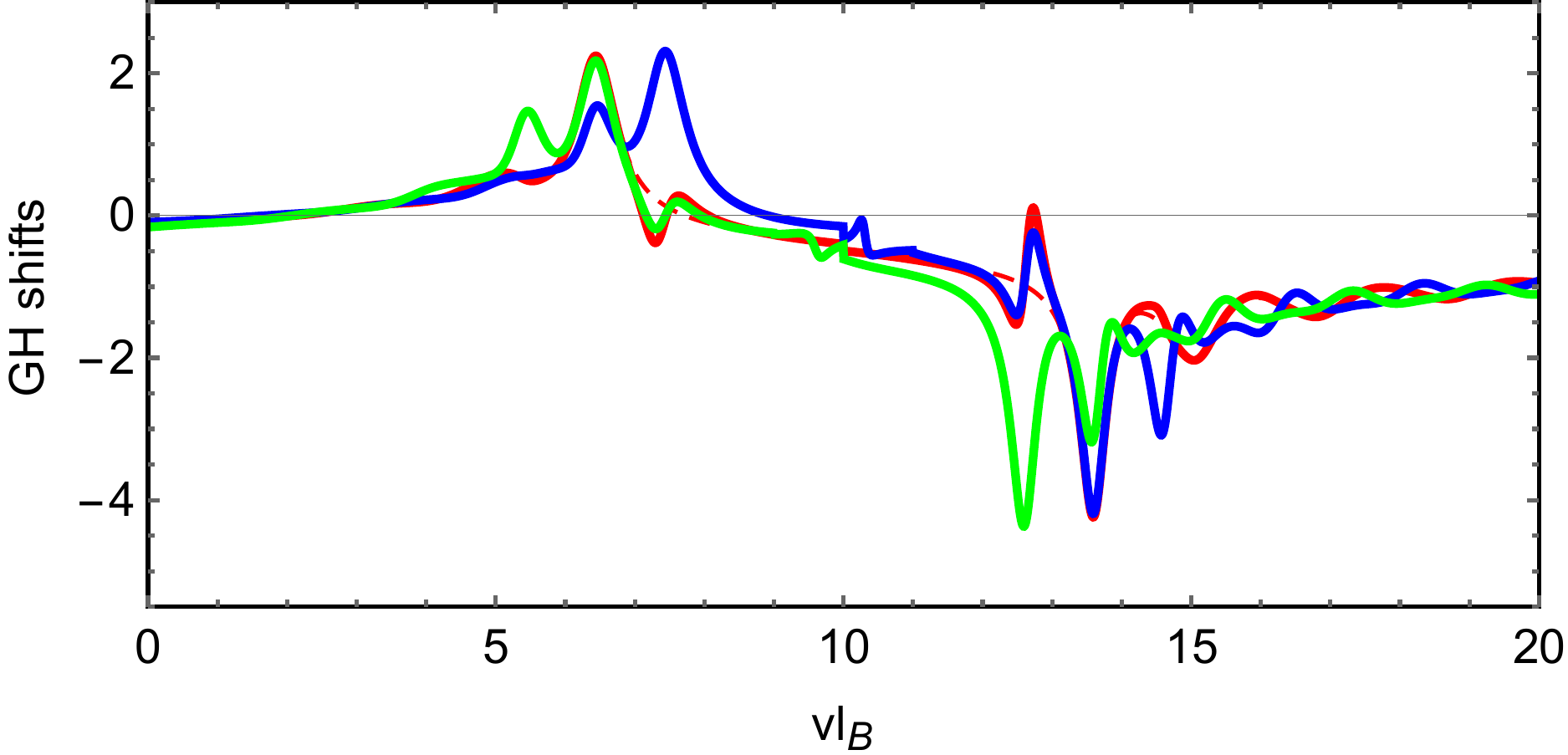}
            \label{db.3}
        }\subfloat[]{
            \centering
            \includegraphics[height=4cm]{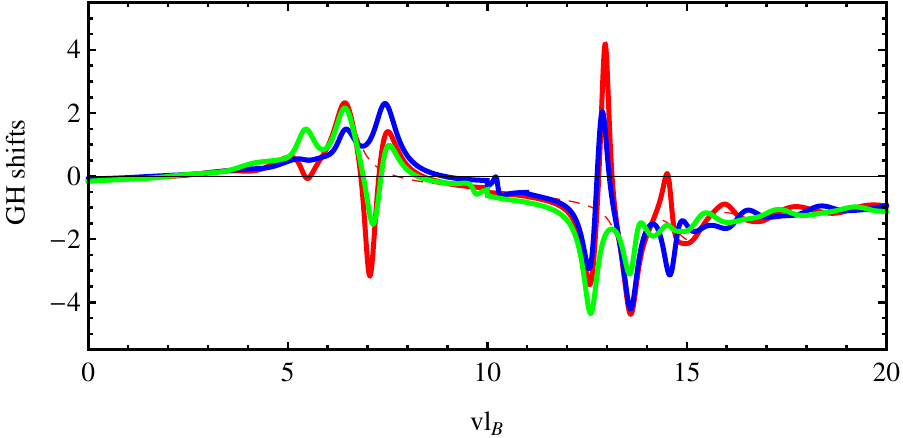}
            \label{db.4}}\\
            \subfloat[]{
            \centering
            \includegraphics[height=4cm]{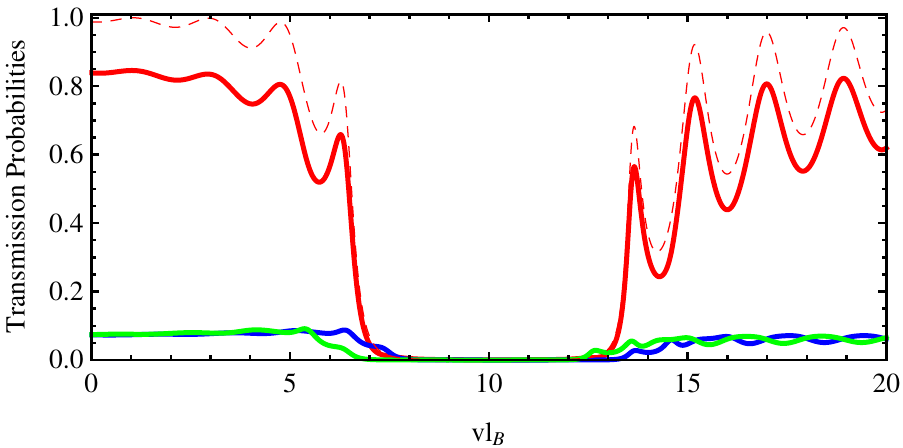}
            \label{db.3}
        }\subfloat[]{
            \centering
            \includegraphics[height=4cm]{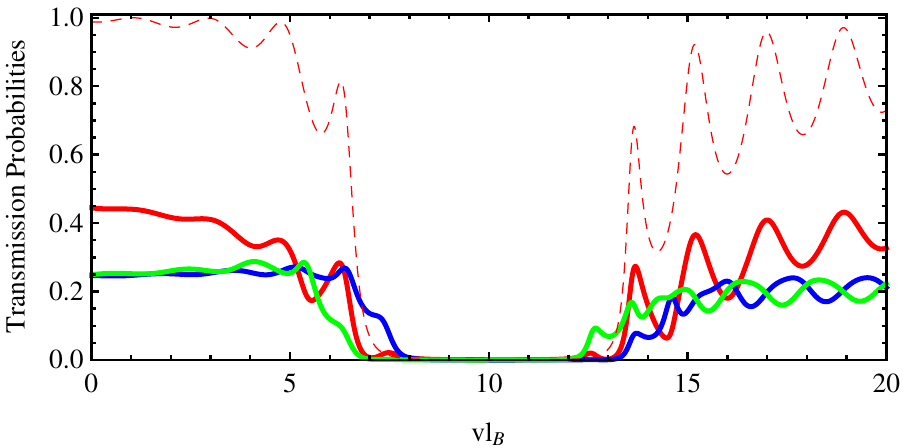}
            \label{db.4}}
    \caption{\sf{(color online) The GH shifts in transmission and transmission
probabilities for central band and first tow sidebands
as a function of the potential $vl_B$ 
for
$\frac{u_j}{\tilde\om}\neq 0$ (oscillating barrier) along with that for static
barrier $\frac{u_j}{\tilde\om}=0$,
 $k_{y} l_{B}=2$, $\epsilon l_{B}=10$, $\frac{L}{l_{B}}=1.5$, $\tilde{\omega} l_{B}=1$, $\tilde{\Delta} l_{B}=1$.
 {\color{blue}(a)/(c)}: $\frac{u_j}{\tilde\om}=0.4$ and
  {\color{blue}(b)/(d)}: $\frac{u_j}{\tilde\om}=0.85$. For no oscillating barrier $(T_{0}, S_{0}^{t})$  (red dashed) and for oscillating
barrier $(T_{0}, S_{0}^{t})$ (red line), $(T_{-1}, S_{-1}^{t})$
(green line), $(T_{1}, S_{1}^{t})$ (blue line). }}
    \label{fig7}
\end{figure}

In Figure \text{\ref{fig7}}, we present  the transmission probabilities $T_l$ and the GH shifts in transmission
$S_{l}^{t}$ for the central band and first two
sidebands 
as a function of the potential $vl_B$
for $\frac{u_j}{\tilde\om}\neq 0$ (oscillating barrier) along with that
for the static barrier $\frac{u_j}{\tilde\om}=0$ 
and 
 specific values $k_{y}l_B=2$, $\epsilon l_B=10$,
$d/l_B = 1.5$, $\tilde{\omega}l_B=1$, $\tilde{\Delta}l_B=1$, such that  $\frac{u_j}{\tilde\om}=0.4$ in Figure
\text{\ref{fig7}}{\color{blue}(a,c)} and $\frac{u_j}{\tilde\om}=0.85$  in Figure \text{\ref{fig7}}{\color{blue}(b,d)}.
Both quantities are showing a series of peaks and resonances where the
resonances correspond to the bound states of the magnetic barrier
for $\frac{u_j}{\tilde\om}=0$ and the oscillating magnetic barrier for $\frac{u_j}{\tilde\om}\neq 0$. We notice that the GH shifts in transmission peak at each bound state energy are clearly shown in the transmission curve underneath. The
energies at which transmission vanishes correspond to energies at
which the GH shifts in transmission change sign. Since these
resonances are very sharp (true bound states with zero width) it
is numerically very difficult to track all of them, if we do this then
the alternation in sign of the GH shifts will be observed. We
notice that around the Dirac point $vl_B=\epsilon l_B+l\tilde{\omega}l_B$ the
number of peaks is equal to the number of transmission resonances. At such a point $T_l$ is showing transmission probabilities for the
central band and the first two sidebands while it oscillates away from
the critical point. We notice that for large values of $vl_B$, the GH
shifts can be positive as well as negative. We deduce that there is a strong dependence of the GH
shifts on the potential height $vl_B$, which can help to realize a
controllable sign of the GH shifts.  We also find that the quantity $\frac{u_j}{\tilde\om}$ is very significant in
determining the GH shifts and transmission probabilities for various
sidebands as shown here. 
This is to be expected as
the probabilities are now spread over the central band and
sidebands. In addition, the maximum transmission through the
oscillating barrier depends on the value of $\frac{u_j}{\tilde\om}$.

\section{Conclusion}

We have studied the effect of both time-oscillating scalar potential
and magnetic field on the Goos-H\"anchen (GH) shifts for the particle transport in gapped graphene. The solutions of energy spectrum are obtained in terms of the physical parameters. We have shown that  
the time-dependent oscillating barrier height generates additional sidebands at energies $\epsilon+l
\tilde{\omega}$ in the transmission probability due to photon absorption or emission.
We have observed that perfect transmission probability at normal incidence (Klein tunneling)
persist for a harmonically driven single barrier.

We have also investigated how the GH shifts in transmission are
affected by various parameters such that the incident angle of the
particles,  width and height of the barrier and oscillation
frequency. Thus our numerical results support the assertion that
quantum interference has an important effect on particle tunneling
through a time-dependent graphene-based single barrier. The
tunneling modes inside the gap and the corresponding 
positive and negative GH shifts are further analyzed. The GH
shifts in transmission for central band and two first sidebands
change sign at the Dirac points $vl_B=\epsilon
l_B+l\tilde{\omega}{{l_B}}$. In particular, the GH
shifts change sign at the transmission zero energies and peaks at
each bound state associated with the single barrier. It is
observed that the GH shifts can be enhanced by the presence of
resonant energies in the system when the incident angle is less
than the critical angle associated with total reflection.

\section*{Acknowledgments}

The generous support provided by the Saudi Center for Theoretical
Physics (SCTP) is highly appreciated by all authors.
AJ and HB acknowledge the support of King Fahd University of Petroleum and Minerals
under research group project RG181001. HB also acknowledges useful consultation with Dr. Michael Vogl.

\begin{appendices}
\numberwithin{equation}{section}
\section{}\label{AppendA}

  To explicitly determine the energy spectrum in our theoretical
model, we separately handle each part of the Hamiltonian
\eqref{ham1}. Thus, let us start from the time-independent Dirac
equation in the absence of oscillating potential for the spinor
$\psi(x, y)=(\psi_{+},\psi_{-})^{T}$ at energy $E$, such as
\begin{equation}
H_{\sf I}\psi(x,y,t)= E \psi(x,y,t)
\end{equation}
where
$\psi(x , y, t)= \psi(x, y)e^{-iEt/\hbar}$.
In matrix form, we have 
\begin{equation}\lb{eq9}
\left(%
\begin{array}{cc}
  0 & -i\partial_{x}-\partial_{y}-\frac{ie}{\hbar c}A(x) \\
  -i\partial_{x}+\partial_{y}+\frac{ie}{\hbar c}A(x) & 0 \\
\end{array}%
\right)\left(%
\begin{array}{c}
  \psi_{+} \\
  \psi_{-} \\
\end{array}%
\right)=\frac{E}{\hbar \upsilon_{F}}\left(%
\begin{array}{c}
  \psi_{+} \\
  \psi_{-} \\
\end{array}%
\right).
\end{equation}
Since the transverse momentum $p_{y}$ is conserved, we can write
the wave function in a separable form {$\psi_{\pm}(x,
y)=\varphi_{\pm}(x)e^{ik_{y}y}$}. Thus after rescaling energy
$\epsilon = E/v_{F}$ and using the unit system with $(\hbar= c = e
= 1)$, we obtain the two linear first order differential equations
\begin{eqnarray}
  && \left (-i\partial_{x}-ik_{y}-i A(x) \right)\varphi_{-}=\epsilon \varphi_{+}\\
&&   \left (-i\partial_{x}+ik_{y}+i A(x)
\right)\varphi_{+}=\epsilon \varphi_{-}.
\end{eqnarray}
These can be combined to describe the solution of \eqref{eq9} and
then  consider the incoming particles to be in plane wave states
$\psi_{inc}(x,y,t)$ at energy $\epsilon$ as
\begin{equation}
\psi_{inc}(x,y,t)=
\begin{pmatrix}
1 \\
 \alpha_{0}\end{pmatrix}
 e^{ik_{0}x}e^{ik_{y}y}e^{-iv_{F}\epsilon t}
\end{equation}
such that $\alpha_{0}$ is given by
\begin{equation}
\alpha_{0}=s_{0}\frac{k_{0} +ik_{y}}{\sqrt{k_{0}^{2}
+k_{y}^{2}}}=s_{0} e^{\textbf{\emph{i}}\phi_{0}}
\end{equation}
where $s_{0}=\mbox{sgn}(\epsilon)$, $\phi_{0}$ is the angle that
the incident particles make with the {$x$-direction}, $k_{0}$ and
$k_{y}$ are the $x$ and $y$-components of the  wave
vector, respectively. After rescaling the potentials $V=v_{F}v$,
$U_j=v_{F}u_j $ and frequency $\omega=v_{F}\tilde{\omega}$, we
show that the transmitted and reflected waves have components at
all energies $\epsilon+ l\tilde{\omega}$ $(l=0, \pm 1, \cdots)$.
Indeed the wave functions $\psi_{re}(x,y,t)$ for reflected
electrons are
\begin{equation}
\psi_{re}(x,y,t)=\sum^{+\infty}_{m,l=-\infty} r_{l}\left(
\begin{array}{c}
1 \\
 -\frac{1}{\alpha_{l}}\end{array}\right)e^{-ik_{l} x +ik_{y}
 y} J_{m-l} \left(\frac{{{u_{j}}}}{\tilde{\omega}}\right)\ e^{-iv_{F}(\epsilon+m\tilde{\omega})t}
\end{equation}
and the corresponding energy reads as
\begin{equation}\lb{energy1}
\epsilon+l\tilde{\omega}=s_{l}\sqrt{k^{2}_{l}+k^{2}_{y}}
\end{equation}
where $r_{l}$ is the reflection amplitude and
$J_{m}\left(\frac{u_j}{\tilde{\omega}}\right)$ is the Bessel
function of the first kind. Note that for the modulation amplitude
$u_j = 0$ we have
 $J_{m-l}
\left(0\right)=\delta_{ml}$. We will return to this point once we
talk about the solution in different regions composing our system.
 The parameter $\al_l$ is the complex number
\begin{equation}
\alpha_{l}=s_{l}\frac{k_{l} +ik_{y}}{\sqrt{k^{2}_{l}
+k_{y}^{2}}}=s_{l}\  e^{\textbf{\emph{i}}\phi_{l}}
\end{equation}
where $\phi_{l}=\tan^{-1}(k_{y}/k_{l})$,
$s_{l}=\mbox{sgn}(\epsilon+l \tilde{\omega})$, the sign again
refers to conduction and valence bands regions. The (number)
wavevector $k_l$ for mode $l$ can be obtained from \eqref{energy1}
to end up with
\begin{equation}
k_{l}=s_{l}\sqrt{\left(\epsilon+l\tilde{\omega}\right)^{2}-k^{2}_{y}}.
\end{equation}
While, the wave functions $\psi_{tr}(x,y,t)$ for transmitted
electrons read as
\begin{equation}
\psi_{tr}(x,y,t)=\sum^{+\infty}_{m,l=-\infty}t_{l}\left(
\begin{array}{c}
1 \\
 \beta_{l}\end{array}\right)e^{ik^{'}_{l} x +ik_{y}
 y}J_{m-l} \left(\frac{{{u_{j}}}}{\tilde{\omega}}\right) e^{-iv_{F}(\epsilon+m\tilde{\omega})t}
\end{equation}
and the eigenvalues
\begin{equation}
\epsilon+l\tilde{\omega}=s_{l}
\sqrt{k^{'2}_{l}+\left(k_{y}+\frac{{{L}}}{l^{2}_{B}}\right)^{2}}
\end{equation}
where the magnetic length $l_{B}=\sqrt{1/B_{0}}$, transmission
amplitude $t_{l}$ and the next complex number
\begin{eqnarray}
&& \beta_{l}=s_{l}\frac{k^{'}_{l}
+i\left(k_{y}+\frac{{{L}}}{l^{2}_{B}}\right)}{\sqrt{k^{'2}_{l}
+\left(k_{y}+\frac{{{L}}}{l^{2}_{B}}\right)^{2}}}=s_{l}\
e^{\textbf{\emph{i}}\theta_{l}}\\
&&
k^{'}_{l}=s_{l}\sqrt{\left(\epsilon+l\tilde{\omega}\right)^{2}-\left(k_{y}+\frac{{{L}}}{l^{2}_{B}}\right)^{2}}\\
&&
\theta_{l}=\tan^{-1}\left[\left(k_{y}+\frac{{{L}}}{l^{2}_{B}}\right)/k^{'}_{l}\right].
\end{eqnarray}

\end{appendices}

\end{document}